\documentclass[a4paper,11pt]{article}
\usepackage{pos}

\usepackage{orcidlink}
\usepackage{physics, mathtools}  
\usepackage{amsmath,amsthm} 
\usepackage{bbold}
\usepackage[english]{babel}

\newcommand{\IdentityMat}{\mathbb{1}}

\renewcommand{\logo}{\relax}

\title{
\textit{Ab initio} Green's functions approach for homogeneous nuclear matter 
}

\author*[a]{F. Marino \orcidlink{0000-0001-7743-1982} }
\author[b,c]{C. Barbieri \orcidlink{0000-0001-8658-6927} }
\author[b,c]{G.~Col\`{o} \orcidlink{0000-0003-0819-1633} }
\author[a,d]{W. G. Jiang \orcidlink{0000-0001-8441-972X} }
\author[e]{S. J. Novario \orcidlink{0000-0001-7836-1269} }

\affiliation[a]{Institut f\"{u}r Kernphysik and PRISMA+ Cluster of Excellence, Johannes Gutenberg-Universit\"{a}t Mainz, 55128 Mainz, Germany}
\affiliation[b]{Dipartimento di Fisica ``Aldo Pontremoli'', Universit\`a degli Studi di Milano, 20133 Milano, Italy}
\affiliation[c]{INFN,  Sezione di Milano, 20133 Milano, Italy}
\affiliation[d]{Mainz Institute for Theoretical Physics, Johannes Gutenberg-Universit\"{a}t, 55128 Mainz, Germany}
\affiliation[e]{Department of Physics, Washington University in Saint Louis, Saint Louis, MO
63130, USA}

\emailAdd{frmarino@uni-mainz.de}

\abstract{
    Homogeneous nuclear matter is investigated using the \textit{ab initio} Self-consistent Green's function (SCGF) approach with nuclear interactions based on chiral effective field theory.
    The employed method, which combines the state-of-the-art algebraic diagrammatic construction approximation at third order with Gorkov correlations, is capable of computing both the equation of state (EOS) and single-particle properties of nuclear matter.
    The EOS calculated with our approach and coupled-cluster theory are shown to agree very well.
    The one-nucleon spectral functions and the momentum distributions are discussed to gain insights into the dynamics of the interacting nuclear matter.
}

\FullConference{10th International Conference on Quarks and Nuclear Physics (QNP2024)\\
8-12 July, 2024\\
Barcelona, Spain\\}

\begin{document}
\maketitle

\section{Introduction}
\label{sec: intro}
Achieving a first-principle description of nuclear phenomena is a long-term endeavor of theoretical nuclear physics.  
At the most fundamental level, nucleons are bound states of quarks, and thus in principle nuclear physics could be described by quantum chromodynamics (QCD) in terms of quarks and gluons. 
However, at present, 
chiral effective field theory ($\chi$EFT) is the widely used framework for understanding nuclear interactions.
$\chi$EFT describes nuclear forces consistently with the symmetries of the underlying theory of QCD, but in terms of the emergent degrees of freedom, i.e. nucleons and pions~\cite{Machleidt2016,computational_nuclear,Hergert2020,EkstromAbInitio}.
Advanced quantum-mechanical methods to solve the many-nucleon problem are the second, and equally crucial, pillar of the \textit{ab initio} approach to nuclear theory~\cite{computational_nuclear,Hergert2020,EkstromAbInitio}.
Only the combination of both frameworks enables to 
provide first-principle predictions with controlled estimates of the theoretical error~\cite{EkstromAbInitio,Hergert2020}.

In this contribution, a study of infinite nuclear matter with the Self-consistent Green's function (SCGF) many-body method~\cite{Barbieri2017,Rios2020,MarinoPhdThesis,Marino2024,Marino2024Theory} is presented.
Nuclear matter is a homogeneous and extended system of interacting nucleons, 
which constitutes an essential model for neutron stars and a key microscopic input to astrophysical simulations~\cite{Burgio2021,HaenselNeutronStars}. 
Therefore, it is crucial to be able to accurately predict nuclear matter properties.
Also, the equation of state (EOS) (i.e, at zero-temperature, the energy per nucleon) is very sensitive to the low-energy constants on which chiral force models depend~\cite{DeltaGo2020,Drischler2021Review}. 

In Sec.~\ref{sec: methods} we introduce the SCGF approach following~\cite{Barbieri2017,MarinoPhdThesis,Marino2024Theory,Marino2024}. 
Predictions for the EOS in comparison with coupled-cluster theory~\cite{Hagen2014,Hagen2014Review}, the single-particle (s.p.) spectral functions (SF), and the momentum distributions are then reported in Sec.~\ref{sec: results}. Perspectives are outlined in Sec.~\ref{sec: conclusions}.

\section{Methods}
\label{sec: methods}
In the SCGF method~\cite{Barbieri2017,Soma2020,Rios2020,Soma2011,Barbieri2022Gorkov}, the central object is the one-body (1B) Green's function (GF), which is determined by solving the Dyson or Gorkov equations, and whose knowledge allows to compute the total ground state (g.s.) energy and the g.s. expectation values of all 1B observables,
While this approach is formally exact, the self-energy $\Sigma^{\star}(\omega)$ entering the Dyson equation must be approximated, in practice, by an expansion in terms of Feynman diagrams. 

Our SCGF variant, see Refs.~\cite{MarinoPhdThesis,Marino2024,Marino2024Theory}, is rooted in the Gorkov framework~\cite{Soma2011,Soma2020,Barbieri2022Gorkov}, and its key equation reads:
\begin{align}
    \label{eq: Gorkov energy dependent}
    \begin{pmatrix}
        T - \mu \IdentityMat + \Sigma^{11}(\omega) & \Sigma^{12(\infty)} \\
        ( \Sigma^{12(\infty)})^{\dagger} & - (T - \mu \IdentityMat ) + \Sigma^{22}(\omega)
    \end{pmatrix}
    \eval_{\omega = \omega_q}
    \begin{pmatrix}
        \mathcal{U}^{q} \\ \mathcal{V}^{q}
    \end{pmatrix} 
    = \hbar \omega_q
    \begin{pmatrix}
        \mathcal{U}^{q} \\ \mathcal{V}^{q}
    \end{pmatrix} .
\end{align}
The unknowns are the excitation energies $\hbar\omega_q$ and amplitudes $\mathcal{U}^{q}$, $\mathcal{V}^{q}$ (for each eigensolution $q$), which fully determine the GF $g^{11}(\omega)$.
The chemical potential $\mu$ must be tuned for each fermion species to ensure the correct number of particles on average.
Due to translational invariance, the GF of homogeneous matter is diagonal in the momentum basis and reads 
\begin{align}
    g^{11}_{\alpha}(\omega) =
    \sum_{q} \frac{ \abs{ \mathcal{V}^{q}_{\alpha}}^2 }{ \hbar\omega - \hbar\omega_q - i\eta} 
    + \frac{ \abs{\mathcal{U}^{q}_{\alpha}}^2 }{ \hbar\omega + \hbar\omega_q + i\eta}, 
\end{align}
with $ \sum_q \abs{ \mathcal{V}^{q}_{\alpha}}^2  + \abs{ \mathcal{U}^{q}_{\alpha}}^2 = 1$. Here, $\alpha$ denotes the s.p. state with quantum numbers $\mathbf{k}_\alpha$ (momentum), $t_\alpha$ (isospin z-projection), and $s_\alpha$ (spin z-projection).  
The Gorkov GF is characterized by pairs of poles $\pm \hbar\omega_q$, which correspond to energies $\epsilon_q = \mu \pm \hbar\omega_q$ symmetric w.r.t. the chemical potential, and have weight $\abs{ \mathcal{V}^{q}_{\alpha}}^2$ ($\abs{ \mathcal{U}^{q}_{\alpha}}^2$) below (above) the Fermi surface $\mu$.
The momentum distribution $\rho_\alpha$, which gives the occupation number for the state $\alpha$ in the interacting system, is readily obtained as $ \rho_\alpha = \sum_q \abs{ \mathcal{V}^{q}_{\alpha}}^2$.
Associated with the GF is the normal spectral function $S^{11}$, defined as
\begin{align}
    \label{eq: spectral function 11}
    S^{11}_{\alpha}(\omega) =
    \sum_q
    \left[
    \abs{ \mathcal{V}^{q}_{\alpha}}^2 \delta(\hbar\omega + \hbar\omega_q)
    + \abs{ \mathcal{U}^{q}_{\alpha}}^2 \delta(\hbar\omega - \hbar\omega_q)
    \right].
\end{align}

In our scheme,
most of the correlations are encoded in
the normal self-energy $\Sigma^{11}(\omega)$, which is approximated using the state-of-the-art algebraic diagrammatic construction scheme~\cite{Barbieri2017,Soma2011,Barbieri2022Gorkov} at third order [ADC(3)]. 
ADC is a powerful way of constructing an approximate self-energy that respects the analytical structure of the exact $\Sigma^{*}$
, and automatically includes contributions to all orders in perturbation theory.
Pairing correlations are taken care of by the anomalous component of self-energy $\Sigma^{12(\infty)}$, which here we approximate at first order~\cite{Marino2024Theory,Marino2024}.

Coupled-cluster theory is also a diagrammatic post-Hartree-Fock (HF) expansion method~\cite{Hagen2014,Hagen2014Review,computational_nuclear}. The CC wave function has the form $\ket{\Psi_0} = e^{T} \ket{\Phi_0}$, with $\ket{\Phi_0}$ being a reference state. $T$, named cluster operator, is a superposition of $n$-particle-$n$-hole excitations.
At the doubles level (CCD), $T$ is truncated as
$T \approx \frac{1}{4} \sum_{ abij } t^{ab}_{ij} c_{a}^{\dagger} c_{b}^{\dagger} c_{j} c_{i}$,
with the $2p2h$ amplitudes $t^{ab}_{ij}$ determined by solving the CC equations. 
Accurate g.s. energies are obtained by adding perturbative triples ($3p3h$) corrections to CCD within the CCD(T) approximation~\cite{Hagen2014,Marino2024}.

In addition, CC and ADC have been combined in the so-called ADC(3)-D schme~\cite{Barbieri2017,Marino2024}.
ADC(3)-D consists in inserting the CC amplitudes from a preliminary CCD calculation into the ADC coupling matrices, and its accuracy has been demonstrated in Ref.~\cite{Marino2024}.

\section{Results}
\label{sec: results}
We present results from the SCGF and CC methods with the ADC(3), ADC(3)-D, and CCD(T) truncations (Sec.~\ref{sec: methods}) using the chiral $\Delta \rm{ NNLO_{go}(450) }$ interaction from Ref.~\cite{DeltaGo2020}.
Both CC and SCGF simulate the homogenous system using a cubic unit cell with a finite number of nucleons~\cite{Marino2023,Marino2024,Hagen2014}, Here, we use $N=66$ neutrons in pure neutron matter (PNM) and $A=132$ 
nucleons in symmetric nuclear matter (SNM)~\cite{Marino2023}. 
Periodic boundary conditions (PBCs) typically suffice to obtain the EOS.
For s.p. properties, twist-averaged boundary conditions (TABCs)~\cite{Hagen2014,Marino2024Theory}, even though more demanding computationally, are preferred, as a denser mesh of $k$-points is obtained, thus allowing
to better approximate the "true" infinite system.

\begin{figure}[h!]
    \centering
    \includegraphics[width=0.4\columnwidth]{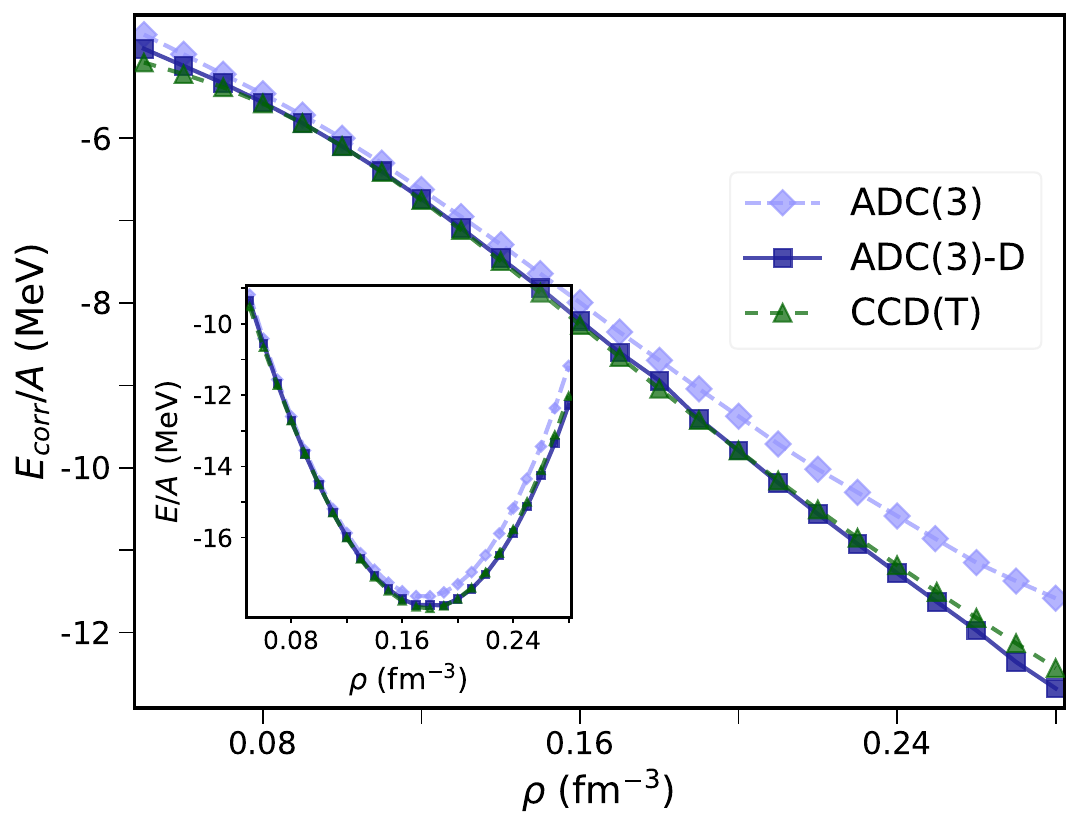}
    \includegraphics[width=0.4\columnwidth]{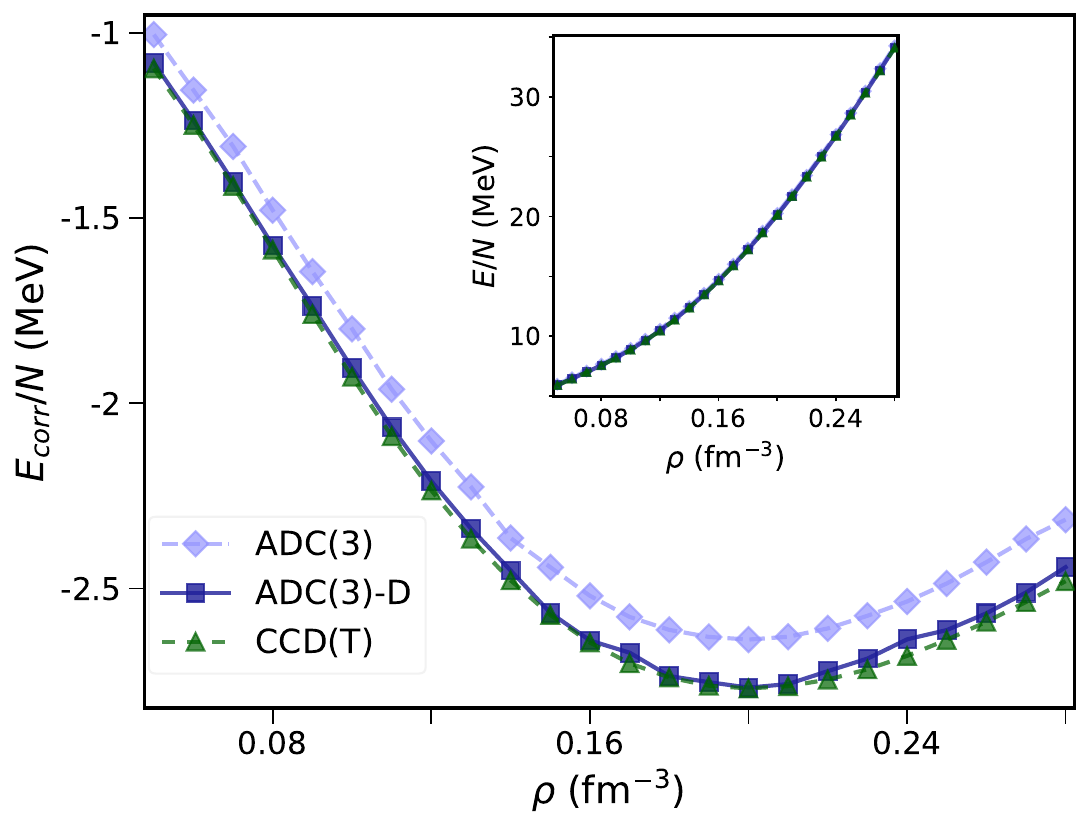}
    \caption{
    Correlation energy per particle (main panel) and energy per particle (inset) as a function of the number density in SNM (left) and PNM (right).
    Calculations have been performed with the ADC(3), ADC(3)-D and CCD(T) methods.
    The $\Delta \rm{ NNLO_{go}(450) }$ interaction from Ref.~\cite{DeltaGo2020} is employed, and PBCs are imposed.
    }
    \label{fig: eos go450}
\end{figure}

The EOS of SNM (left) and PNM (right) are reported in Fig.~\ref{fig: eos go450}. 
The inset shows the energies per particle, and an excellent agreement between the three sets of computations is found. 
The main panels investigate the correlation energy per particle, and we can appreciate that predictions are very similar also on this scale.
This confirms that these sophisticated many-body methods describe nuclear matter accurately~\cite{Marino2024}.
Note also that the inclusion of CC-corrected coupling vertices in ADC(3)-D leads to very close agreement with CCD(T)~\cite{Marino2024}.

\begin{figure}[h!]
    \centering
    \includegraphics[width=0.45\columnwidth]{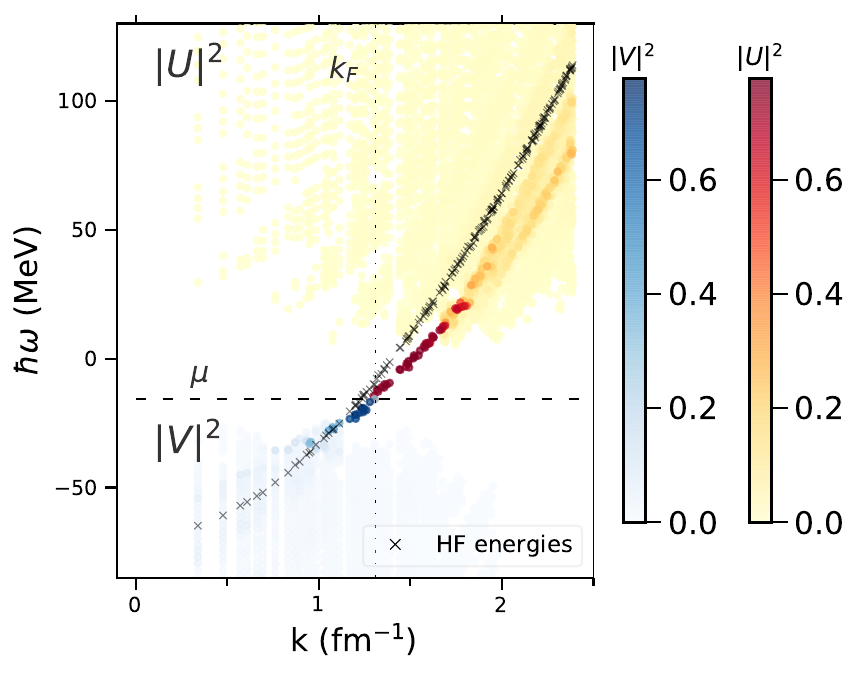}
    \includegraphics[width=0.45\columnwidth]{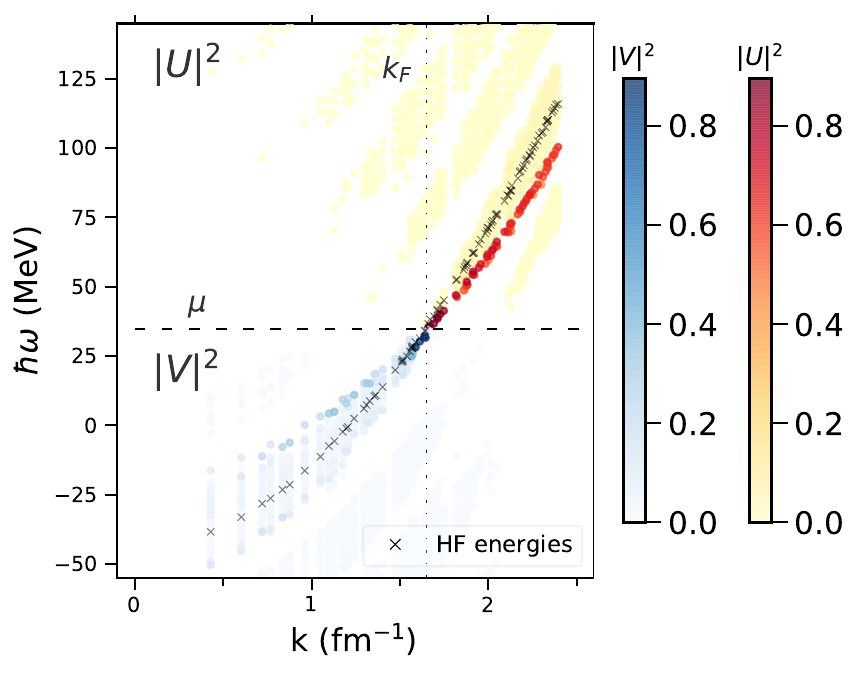}
    \caption{
    Two-dimensional representation of the spectral function for SNM (left) and PNM (right) as a function of the momentum $k$ and the energy $\hbar\omega$.
    Calculations are performed at $\rho=0.16\,\rm{fm}^{-3}$ with ADC(3) employing TABCs with the $\Delta \rm{ NNLO_{go}(450) }$ interaction.
    The Fermi momentum $k_F$ and the chemical potential $\mu$ are marked by dotted and dashed lines.
    The squared amplitudes $\abs{\mathcal{U}^{q}_{\alpha}}^2$ ($\abs{ \mathcal{V}^{q}_{\alpha}}^2$) are shown for poles above (below) the Fermi level. 
    The color scale is shown next to each plot. HF s.p. energies are represented by crosses.
    For SNM, the neutron part of the SF is displayed.
    }
    \label{fig: snm 2d spectral}
\end{figure}

In Fig.~\ref{fig: snm 2d spectral}, the SF~\eqref{eq: spectral function 11} is shown for SNM (left) and PNM (right).
$S^{11}$ is represented as a two-dimensional map as a function of the momentum $k = \abs{ \mathbf{k}_{\alpha} }$ (horizontal axis) and the energy $\hbar\omega$ (vertical axis).
Poles $\epsilon_q$ are denoted by dots, and a color scale is used to represent the value of the strengths $\abs{\mathcal{V}^{q}_{\alpha}}^2$ ($\abs{ \mathcal{U}^{q}_{\alpha}}^2$) for energies below (above) the Fermi level $\mu$. 
The strong correlations that characterize SNM are manifest in the increased fragmentation of the SF at momenta $k>1.5\,\rm{fm}^{-1}$. 
In both SNM and PNM, the dominant peaks, as well as HF energies (crosses), follow a roughly parabolic trend as a function of $k$, which is driven by the kinetic energy $\sim k^2$.
The SNM spectrum is highly fragmented and shows a wealth of satellite peaks, especially for large $k$ and $\hbar\omega$.
In PNM, where correlations are weaker, the background is fainter and the primary branch dominates well into the high-momentum region.
In both cases, states close to the Fermi momentum 
feature a single dominant pole with $\abs{\mathcal{V}^{q}_{\alpha}}^2 \simeq 1$ for $k < k_F$ or $\abs{\mathcal{U}^{q}_{\alpha}}^2 \simeq 1$ for $k > k_F$.
Our \textit{ab initio} calculations thus validate microscopically the picture of
Landau quasi-particle excitations at the Fermi surface~\cite{Rios2012,DickhofVanNeck}.

\begin{figure}[h!]
    \centering
    \includegraphics[width=0.4\columnwidth]{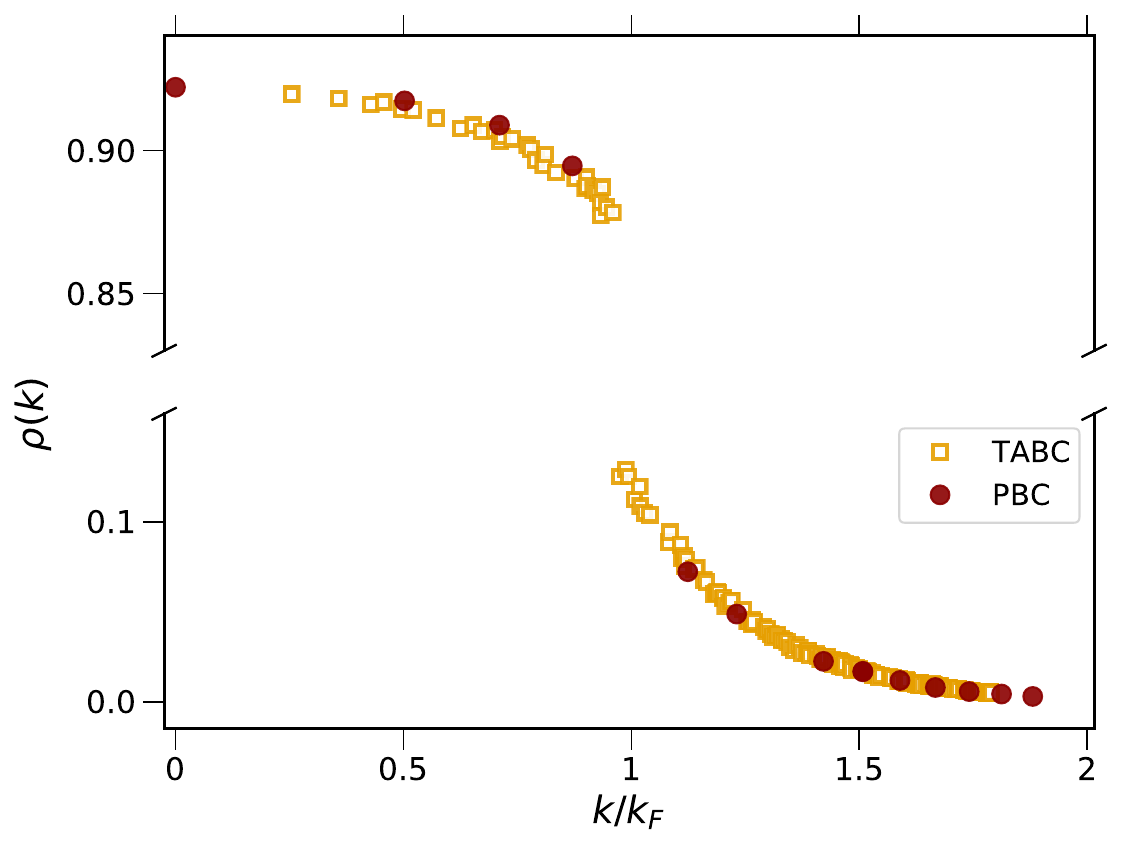}
    \includegraphics[width=0.4\columnwidth]{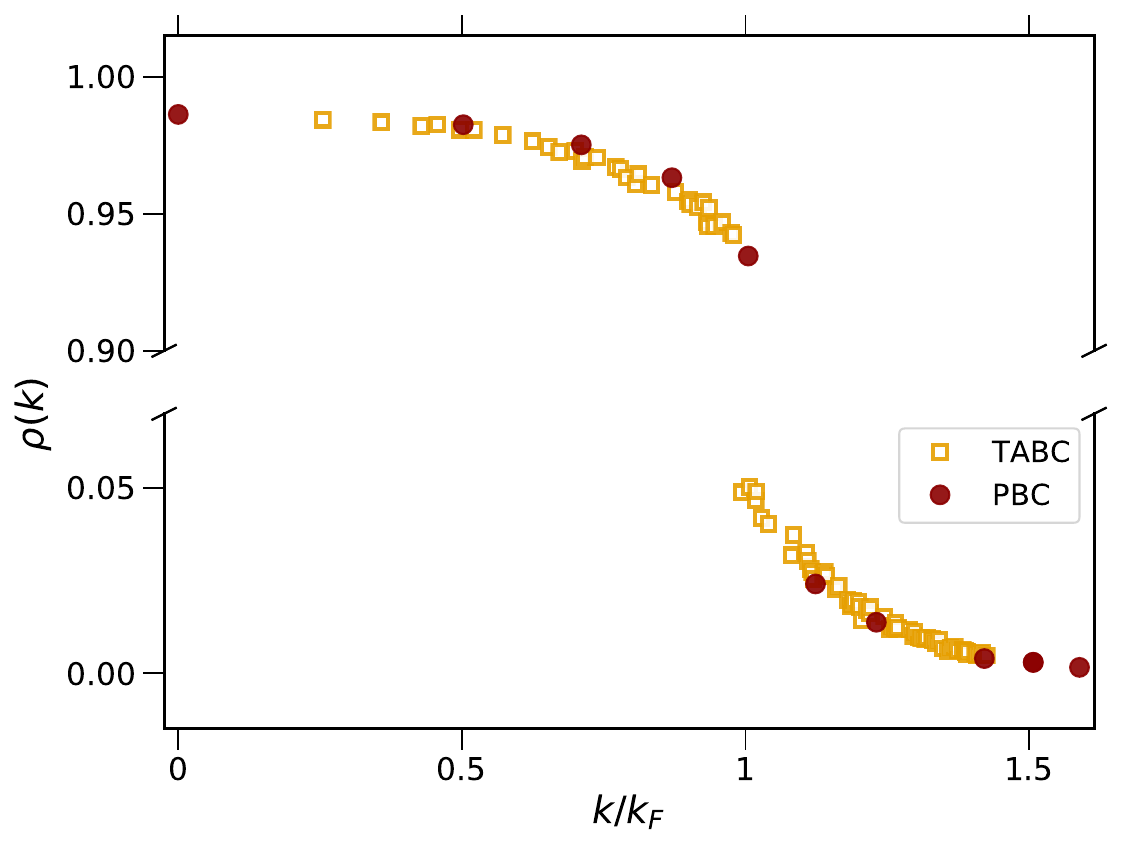}
    \caption{
    Momentum distributions $\rho(k)$ as a function of $k/k_F$ for SNM (left) and PNM (right) at density $\rho=0.16\,\rm{fm}^{-3}$. Calculations performed with ADC(3) with PBC (circles) and TABC (empty squares) are compared.
    Note that different scales are used on the vertical axis for points below and above the discontinuity.
    }
    \label{fig: momentum distr}
\end{figure}

Finally, momentum distributions $ \rho(k)$  in SNM (left) and PNM (right) with both PBCs (circles) and TABCs (empty squares) are shown in Fig.~\ref{fig: momentum distr}.
While in a HF picture states below (above) the Fermi momentum are completely occupied (empty), in the interacting system hole states ($k<k_F$) are partially depleted. 
The depletion is much larger in SNM,
due to stronger correlations.
Conversely, a momentum tail appears above $k_F$, with particle states being partially populated, while a finite discontinuity across the Fermi surface is preserved also in the presence of interactions. 

\section{Conclusions and perspectives}
\label{sec: conclusions}
We have investigated infinite nuclear matter using the ADC-SCGF \textit{ab initio} method.
We have studied both the EOS, which agrees very well with coupled-cluster~\cite{Marino2024}, and single-particle properties.
We plan to apply ADC to superfluid neutron matter, quasi-particle properties (effective mass, lifetimes), and momentum distributions across densities and isospin asymmetries.

\bibliographystyle{JHEP}
\bibliography{bibliography.bib} 

\end{document}